\documentclass[fleqn,11pt]{article}

\usepackage{latexsym,ifthen,epsfig,color}
\usepackage{amsmath,amssymb,amsthm}
\usepackage{float}

\newcommand{\join}{\text{\textcircled{{\footnotesize 1}}}}
\newcommand{\cojoin}{\text{\textcircled{{\footnotesize 0}}}}

\newcommand{\NP}{\ensuremath{\mathbb{NP}}}

\newtheorem{theorem}{Theorem}
\newtheorem{lemma}{Lemma}
\newtheorem{corollary}{Corollary}

\newtheorem{clai}{Claim}
\newtheorem{observation}{Observation}

\begin{document}

\title{Finding Dominating Induced Matchings in $P_8$-Free Graphs in Polynomial Time}

\author{
Andreas Brandst\"adt\footnote{Fachbereich Informatik,
Universit\"at Rostock, A.-Einstein-Str. 21, D-18051 Rostock, Germany,
{\texttt ab@informatik.uni-rostock.de}}
\and
Raffaele Mosca\footnote{Dipartimento di Economia, Universit\'a degli Studi ``G. D'Annunzio''
Pescara 65121, Italy.
{\texttt r.mosca@unich.it}}
}

\maketitle

\begin{abstract}
Let $G=(V,E)$ be a finite undirected graph. An edge set $E' \subseteq E$ is a {\em dominating induced matching} ({\em d.i.m.}) in $G$ if every edge in $E$ is intersected by exactly one edge of $E'$.
The \emph{Dominating Induced Matching} (\emph{DIM}) problem asks for the existence of a d.i.m.\ in $G$; this problem is also known as the \emph{Efficient Edge Domination} problem.

The DIM problem is related to parallel resource allocation problems, encoding theory and network routing. It is \NP-complete even for very restricted graph classes such as planar bipartite graphs with maximum degree three and is solvable in linear time for $P_7$-free graphs. However, its complexity was open for $P_k$-free graphs for any $k \ge 8$; $P_k$ denotes the chordless path with $k$ vertices and $k-1$ edges. We show in this paper that the weighted DIM problem is solvable in polynomial time for $P_8$-free graphs.
\end{abstract}


\section{Introduction}\label{sec:intro}

Let $G=(V,E)$ be a finite undirected graph. A vertex $v \in V$ {\em dominates} itself and its neighbors. A vertex subset $D \subseteq V$ is an {\em efficient dominating set} ({\em e.d.s.} for short) of $G$ if every vertex of $G$ is dominated by exactly one vertex in $D$.
The notion of efficient domination was introduced by Biggs \cite{Biggs1973} under the name {\em perfect code}.
The {\sc Efficient Domination} (ED) problem asks for the existence of an e.d.s.\ in a given graph $G$ (note that not every graph has an e.d.s.)

If a vertex weight function $\omega: V \to \mathbb{N}$ is given, the {\sc Weighted Efficient Domination} (WED) problem asks for a minimum weight e.d.s.\ in $G$, if there is one, or for determining that $G$ has no e.d.s.

A set $M$ of edges in a graph $G$ is an \emph{efficient edge dominating set} (\emph{e.e.d.s.} for short) of $G$ if and only if it is an e.d.s.\ in its line graph $L(G)$. The {\sc Efficient Edge Domination} (EED) problem asks for the existence of an e.e.d.s.\ in a given graph $G$. Thus, the EED problem for a graph $G$ corresponds to the ED problem for its line graph $L(G)$. Again, note that not every graph has an e.e.d.s. An efficient edge dominating set is also called \emph{dominating induced matching} ({\em d.i.m}.\ for short) and the EED problem is called the {\sc Dominating Induced Matching} (DIM) problem in some papers (see e.g. \cite{BraHunNev2010,BraMos2014,CarKorLoz2011}); subsequently, we will use this notation in the manuscript. The edge-weighted version of DIM for graph $G$ corresponds to the vertex-weighted version of ED for $L(G)$.

In \cite{GriSlaSheHol1993}, it was shown that the DIM problem is \NP-complete; see also~\cite{BraHunNev2010,CarKorLoz2011,LuKoTan2002,LuTan1998}.
However, for various graph classes, DIM is solvable in polynomial time. For mentioning some examples, we need the following notions: 

Let $P_k$ denote the chordless path $P$ with $k$ vertices, say $a_1,\ldots,a_k$, and $k-1$ edges $a_ia_{i+1}$, $1 \le i \le k-1$; we also denote it as $P=(a_1,\ldots,a_k)$. 
 
For indices $i,j,k \ge 0$, let $S_{i,j,k}$ denote the graph with vertices $u,x_1,\ldots,x_i$, $y_1,\ldots,y_j$, $z_1,\ldots,z_k$ such that the subgraph induced by $u,x_1,\ldots,x_i$ forms a $P_{i+1}$ $(u,x_1,\ldots,x_i)$, the subgraph induced by $u,y_1,\ldots,y_j$ forms a $P_{j+1}$ $(u,y_1,\ldots,y_j)$, and the subgraph induced by $u,z_1,\ldots,z_k$ forms a $P_{k+1}$ $(u,z_1,\ldots,z_k)$, and there are no other edges in $S_{i,j,k}$. Thus, {\em claw} is $S_{1,1,1}$, and $P_k$ is isomorphic to e.g. $S_{0,0,k-1}$.

DIM is solvable in polynomial time for $S_{1,1,1}$-free graphs \cite{CarKorLoz2011}, for $S_{1,2,3}$-free graphs \cite{KorLozPur2014}, and 
for $S_{2,2,2}$-free graphs \cite{HerLozRieZamdeW2015}. In \cite{HerLozRieZamdeW2015}, it is conjectured that for every fixed $i,j,k$, DIM is solvable in polynomial time for $S_{i,j,k}$-free graphs (actually, an even stronger conjecture is mentioned in \cite{HerLozRieZamdeW2015}); this includes $P_k$-free graphs for $k \ge 8$.
In \cite{BraMos2014}, DIM is solved in linear time for $P_7$-free graphs. 

\medskip

In this paper we show that edge-weighted DIM can be solved in polynomial time for $P_8$-free graphs.

\section{Definitions and Basic Properties}\label{sec:basicnotionsresults}

\subsection{Basic notions}\label{subsec:basicnotions}

Let $G$ be a finite undirected graph without loops and multiple edges. Let $V$ denote its vertex set and $E$ its edge set; let $|V|=n$ and $|E|=m$.
For $v \in V$, let $N(v):=\{u \in V \mid uv \in E\}$ denote the {\em open neighborhood of $v$}, and let $N[v]:=N(v) \cup \{v\}$ denote the {\em closed neighborhood of $v$}. If $xy \in E$, we also say that $x$ and $y$ {\em see each other}, and if  $xy \not\in E$, we say that $x$ and $y$ {\em miss each other}. A vertex set $S$ is {\em independent} in $G$ if for every pair of vertices $x,y \in S$, $xy \not\in E$. A vertex set $Q$ is a {\em clique} in $G$ if for every pair of vertices $x,y \in Q$, $x \neq y$, $xy \in E$. For $uv \in E$ let $N(uv):= N(u) \cup N(v) \setminus \{u,v\}$ and $N[uv]:= N[u] \cup N[v]$.

For $U \subseteq V$, let $G[U]$ denote the subgraph of $G$ induced by vertex set $U$. Clearly $xy \in E$ is an edge in $G[U]$ exactly when $x \in U$ and $y \in U$; thus, $G[U]$ will be often denoted simply by $U$ when that is clear in the context.

Let $A$ and $B$ be disjoint sets of vertices of $G$. If a vertex from $A$ sees a vertex from $B$, we say that {\em $A$ and $B$ see each other}.
If every vertex from $A$ sees every vertex from $B$ then we denote this by $A \join B$. In particular, if a vertex $u \notin B$ sees all vertices of $B$ then we denote this by $u \join B$ (in this case, $u$ is called {\em universal} for $B$).    
If every vertex from $A$ misses every vertex from $B$, we say that {\em $A$ and $B$ miss each other} and denote this by $A \cojoin B$. If for $A' \subseteq A$, $A' \cojoin (A \setminus A')$ holds, we say that $A'$ is {\em isolated} in $A$.

As already mentioned, a {\em chordless path} $P_k$ has $k$ vertices, say $v_1,\ldots,v_k$, and edges $v_iv_{i+1}$, $1 \le i \le k-1$. The {\em length} of $P_k$ is $k-1$.
A {\em chordless cycle} $C_k$ has $k$ vertices, say $v_1,\ldots,v_k$, and edges $v_iv_{i+1}$, $1 \le i \le k-1$, and $v_kv_1$. The {\em length} of $C_k$ is $k$.
 
Let $K_i$ denote the clique with $i$ vertices. Let $K_4-e$ or {\em diamond} be the graph with four vertices and five edges, say vertices $a,b,c,d$ and edges $ab,ac,bc,bd,cd$; its {\em mid-edge} is the edge $bc$. 
A {\em gem} has five vertices, say, $a,b,c,d,e$, such that $(a,b,c,d)$ forms a $P_4$ and $e$ is universal for $\{a,b,c,d\}$. 
A {\em butterfly} has five vertices and six edges, say, $a,b,c,d,e$ and edges $ab,ac,bc,cd,ce,de$. The {\em peripheral edges} of the butterfly are $ab$ and $de$.
A {\em star} is a graph formed by an independent set $I$ plus one vertex (the {\em center} of the star) which is universal for $I$; in particular let us say that a star is {\em trivial} if it is an edge or a single vertex, and is {\em non-trivial} otherwise.

\medskip

We often consider an edge $e = uv$ to be a set of two vertices; then it makes sense to say, for example, $u \in e$ and $e \cap e' \neq \emptyset$ for an edge $e'$. For two vertices $x,y \in V$, let $dist_G(x,y)$ denote the {\em distance between $x$ and $y$ in $G$}, i.e., the length of a shortest path between $x$ and $y$ in $G$. The {\em distance between two edges} $e,e' \in E$ is the length of a shortest path between $e$ and $e'$, i.e., $dist_G(e,e')= \min\{dist_G(u,v) \mid u \in e, v \in e'\}$. In particular, this means that $dist_G(e,e')=0$ if and only if $e \cap e' \neq \emptyset$. 

An edge set $M \subseteq E$ is an {\em induced matching} if its members have pairwise distance at least~2. 
Obviously, if $M$ is a d.i.m.\ then $M$ is an induced matching.

\medskip

For an edge $xy$, let $N_i(xy)$ denote the {\em distance levels of $xy$}:
$$N_i(xy):=\{z \in V \mid dist_G(z,xy) = i\}.$$

For a set ${\cal F}$ of graphs, a graph $G$ is called {\em ${\cal F}$-free} if $G$ contains no induced subgraph from ${\cal F}$.
A graph is {\em hole-free} if it is $C_k$-free for all $k \ge 5$.
A graph is {\em weakly chordal} if it is $C_k$-free and $\overline{C_k}$-free for all $k \ge 5$, i.e., the graph and its complement are hole-free.

\medskip

If $M$ is a d.i.m.\ then an edge is {\em matched by $M$} if it is either in $M$ or shares a vertex with some edge in $M$.
Note that $M$ is a d.i.m.\ in $G$ if and only if it corresponds to a dominating set (of vertices) in the line graph $L(G)$ and an independent set of vertices in the square $L(G)^2$. The {\sc Maximum Weight Independent Set} (MWIS) problem asks for a maximum weight independent set in a given graph with vertex-weight function.
The DIM problem for $G$ can be reduced to the MWIS problem for $L(G)^2$ (see \cite{BraLeiRau2012}). For instance, in \cite{CamSriTan2003}, it is shown that for weakly chordal graphs $G$, $L(G)^2$ is weakly chordal, and since MWIS can be solved in polynomial time for weakly chordal graphs \cite{SpiSri1995}, DIM can be solved in polynomial time for weakly chordal graphs as well. Actually, DIM can be solved in polynomial time even for hole-free graphs \cite{BraHunNev2010}.

\medskip

$P_8$-free graphs having a d.i.m.\ are $C_k$-free for $k \ge 9$ and $\overline{C_k}$-free for $k \ge 6$ (see Corollary~\ref{cly:k4gemfree} below) but we do not yet have a proof that, using the reduction to $L(G^2)$, DIM can be solved in polynomial time for $P_8$-free graphs; our approach in this paper is a direct one following the approach for $P_7$-free graphs given in \cite{BraMos2014}.

\subsection{Forbidden subgraphs and forced edges}

The subsequent observations are helpful (some of them are mentioned e.g. in \cite{BraHunNev2010,BraMos2014}); since we deal with the larger class of $P_8$-free graphs instead of $P_7$-free graphs and in order to make this manuscript self-contained, we give all proofs where forbidding $P_8$ plays a role.

\begin{observation}[\cite{BraHunNev2010,BraMos2014}]\label{dimC3C5C7C4}
Let $M$ be a d.i.m.\ in $G$.
\begin{itemize}
\item[$(i)$] $M$ contains at least one edge of every odd cycle $C_{2k+1}$ in $G$, $k \ge 1$,
and exactly one edge of every odd cycle $C_3$, $C_5$, $C_7$ of $G$.
\item[$(ii)$] No edge of any $C_4$ can be in $M$.
\item[$(iii)$] For each $C_6$ either exactly two or none of its edges are in $M$.
\end{itemize}
\end{observation}

{\bf Proof.} See Observation 2 in \cite{BraMos2014}.

\medskip

Since every triangle contains exactly one $M$-edge and no $M$-edge is in any $C_4$, and the pairwise distance of edges in any d.i.m.\ is at least 2, we obtain:

\begin{corollary}\label{cly:k4gemfree}
If a graph $G$ has a d.i.m.\ then $G$ is $K_4$-free, gem-free and $\overline{C_k}$-free for any $k \ge 6$.
\end{corollary}

\medskip

As a consequence of Observation \ref{dimC3C5C7C4} $(ii)$, we give all edges in any $C_4$ of $G$ weight $\infty$. Note that a d.i.m.\ of finite weight cannot contain any edge of a $C_4$.

\medskip

If an edge $e \in E$ is contained in {\em every} d.i.m.\ of $G$, we call it a {\em forced} edge of $G$.
\begin{observation}\label{obs:diamondbutterfly}
The mid-edge of any diamond in $G$ and the two peripheral edges of any induced butterfly are forced edges of $G$.
\end{observation}

Note that in a graph with d.i.m., the set of forced edges is an induced matching. So our algorithm solving the DIM problem on $P_8$-free graphs has to check whether the set of forced edges is an induced matching (and finally might be extended to a d.i.m.\ of $G$). If $M$ is an induced matching of already collected forced edges and edge $vw$ is a new forced edge, we can reduce the graph as follows: 

\medskip

{\bf Reduction-Step-($vw,M$).}
\begin{itemize}
\item[ ]
If $M \cup \{vw\}$ is not an induced matching then STOP - $G$ has no d.i.m., otherwise add $vw$ to $M$, i.e., $M:=M \cup \{vw\}$, delete $v$ and $w$ and all edges incident to $v$ and $w$ in $G$, and give all edges that were at distance 1 from $vw$ in $G$ weight~$\infty$.
\end{itemize}

Obviously, the graph resulting from the reduction step is an induced subgraph of $G$. In particular, edges with weight $\infty$ are not in any d.i.m.\ of finite weight in $G$.

\begin{observation}[\cite{BraMos2014}]\label{obs:redstep}
Let $M'$ be an induced matching which is a set of forced edges in $G$. Then $G$ has a d.i.m.\ $M$ if and only if after applying the reduction step to all edges in $M'$, the resulting graph has a d.i.m.\ $M \setminus M'$.
\end{observation}

Subsequently, this approach will often be used. Note that after applying the Reduction Step to all mid-edges of diamonds and all peripheral edges of butterflies in $G$, the resulting graph is (diamond, butterfly)-free. By Corollary \ref{cly:k4gemfree}, a graph $G$ having a d.i.m.\ is $K_4$-free. 
Thus, from now on, we can assume that $G$ is ($P_8$,$K_4$, diamond, butterfly)-free.

\section{The Structure of $P_8$-Free Graphs With a Dominating Induced Matching}\label{P8free}

Throughout this section, let $G=(V,E)$ be a connected $(P_8$, $K_4$, diamond, butterfly)-free graph having a d.i.m.\ $M$.  Note that if $G$ has a d.i.m.\ $M$ and $V(M)$ denotes the vertex set of $M$ then $V \setminus V(M)$ is an independent set, say $I$, i.e., 
\begin{equation}\label{IV(M)partition}
V \mbox{ has the partition } V = I \cup V(M). 
\end{equation}

\subsection{The distance levels of an $M$-edge $xy$ in a $P_3$}\label{subsec:distlevels}

We first describe some general structure properties for the distance levels of an edge in a d.i.m.
Since $G$ is $(K_4$, diamond, butterfly)-free, we have:
\begin{observation}\label{obse:neighborhood}
For every vertex $v$ of $G$, $N(v)$ is the disjoint union of isolated vertices and at most one edge. Moreover, for every edge $xy \in E$, there is at most one common neighbor of $x$ and $y$.
\end{observation}

Since it is trivial to check whether $G$ has a d.i.m.\ with exactly one edge, from now on we can assume that $|M| \geq 2$. Since $G$ is connected and butterfly-free, we have:
\begin{observation}\label{obse:xy-in-P3}
If $|M| \geq 2$ then there is an edge in $M$ which is contained in a $P_3$ of $G$.
\end{observation}

Let $xy \in M$ be an $M$-edge for which there is a vertex $r$ such that $\{r,x,y\}$ induce a $P_3$ with edge $rx \in E$. We consider a partition into the distance levels $N_i=N_i(xy)$, $i \ge 1$, with respect to the edge $xy$. 
By (\ref{IV(M)partition}) and since we assume that $xy \in M$, clearly, $N_1 \subseteq I$ and thus:
\begin{equation}\label{N1subI}
N_1 \mbox{ is an independent set.}
\end{equation}

Since $G$ is $P_8$-free and $xy$ is contained in a $P_3$ $\{r,x,y\}$ of $G$, we obtain:
\begin{equation}\label{N6empty}
N_k=\emptyset \mbox{ for } k \geq 6.
\end{equation}

\noindent
{\em Proof of} (\ref{N6empty}):
If $N_6 \neq \emptyset$ then there are vertices $v_i \in N_i$, $2 \le i \le 6$, such that $\{v_6,v_5,v_4,v_3,v_2\}$ induce a chordless path with $v_iv_{i+1} \in E$ for $2 \le i \le 5$. 
  If $v_2r \in E$ then $\{v_6,v_5,v_4,v_3,v_2,r,x,y\}$ would induce a $P_8$ in $G$. Thus, $v_2r \notin E$; let $v_1 \in N_1$ be a neighbor of $v_2$. 
By (\ref{N1subI}), $v_1r \notin E$. Now, if $v_1x \in E$ then $\{v_6,v_5,v_4,v_3,v_2,v_1,x,r\}$ induce a $P_8$ in $G$, and if $v_1x \notin E$ then $v_1y \in E$ and thus, $\{v_6,v_5,v_4,v_3,v_2,v_1,y,x\}$ induce a $P_8$ in $G$ which is a contradiction.
\qed

\medskip

Subsequently, the principle of the proof of (\ref{N6empty}) will be applied in various cases whenever a $P_8$ has to be excluded. 

\medskip

Since $xy \in M$, no edge between $N_1$ and $N_2$ is in $M$. Since $N_1 \subseteq I$ and all neighbors of vertices in $I$ are in $V(M)$, we have:
\begin{equation}\label{N2M2S2}
N_2 \mbox{ is the disjoint union of edges and isolated vertices. }
\end{equation}

Let $M_2$ denote the set of edges with both ends in $N_2$ and let $S_2 = \{u_1,\ldots,u_k\}$ denote the set of isolated vertices in $N_2$; $N_2=V(M_2) \cup S_2$ is a partition of $N_2$. Obviously:
\begin{equation}\label{M2subM}
M_2 \subseteq M \mbox{ and } S_2 \subseteq V(M).
\end{equation}

If for $xy \in M$, an edge $e \in E$ is contained in {\em every} dominating induced matching $M$ of $G$ with $xy \in M$, we say that $e$ is an {\em $xy$-forced} $M$-edge. The Reduction Step for forced edges can also be applied for $xy$-forced $M$-edges (then, in the unsuccessful case, $G$ has no d.i.m.\ containing $xy$). 
We do this whenever an $xy$-forced $M$-edge is found. The first example is the following one; obviously, by (\ref{M2subM}), we have:
\begin{equation}\label{M2xymandatory}
\mbox{Every edge in } M_2 \mbox{ is an $xy$-forced $M$-edge}.
\end{equation}

Thus, from now on, we can assume that $M_2=\emptyset$, i.e., $N_2=S_2 = \{u_1,\ldots,u_k\}$. For every $i \in \{1,\ldots,k\}$, let $u'_i \in N_3$ denote the {\em $M$-mate} of $u_i$ (i.e., $u_iu'_i \in M$). Let $M_3=\{u_iu'_i: i \in \{1,\ldots,k\}\}$ denote the set of $M$-edges with one endpoint in $S_2$ (and the other endpoint in $N_3$). Obviously, by (\ref{M2subM}) and the distance condition for a d.i.m.\ $M$, the following holds:
\begin{equation}\label{noMedgesN3N4}
\mbox{ No edge with both ends in } N_3 \mbox{ and no edge between } N_3 \mbox{ and } N_4 \mbox{ is in } M.
\end{equation}

As a consequence of (\ref{noMedgesN3N4}) and the fact that every triangle contains exactly one $M$-edge (see Observation \ref{dimC3C5C7C4} $(i)$), we have:
\begin{equation}\label{triangleaN3bcN4}
\mbox{For every triangle $abc$} \mbox{ with } a \in N_3, \mbox{ and } b,c \in N_4, \mbox{ $bc \in M$ is an $xy$-forced $M$-edge}.
\end{equation}

This means that for the edge $bc$, the Reduction Step can be applied, and from now on, we can assume that there is no such triangle $abc$ with $a \in N_3$ and $b,c \in N_4$, i.e., for every edge $uv \in E$ in $N_4$:
\begin{equation}\label{edgeN4N3neighb}
(N(u) \cap N_3) \cap (N(v) \cap N_3) = \emptyset.
\end{equation}

\medskip

According to $(\ref{M2subM})$ and the assumption that $M_2=\emptyset$ (recall $N_2 = \{u_1,\ldots,u_k\}$), let:
\begin{enumerate}
\item[ ] $T_{one} := \{t \in N_3: |N(t) \cap N_2| = 1\}$;

\item[ ] $T_i := T_{one} \cap N(u_i)$, $i \in \{1,\ldots,k\}$;

\item[ ] $S_3 := N_3 \setminus T_{one}$.
\end{enumerate}

By definition, $T_i$ is the set of {\em private} neighbors of $u_i$ in $N_3$ (note that $u'_i \in T_i$), and
$T_1 \cup \ldots \cup T_k$ is a partition of $T_{one}$, and $T_{one} \cup S_3$ is a partition of~$N_3$.

\begin{lemma}\label{lemm:structure2}
The following statements hold:
\begin{enumerate}
\item[$(i)$] For all $i \in \{1,\ldots,k\}$, $T_i \cap V(M)=\{u_i'\}$.

\item[$(ii)$] For all $i \in \{1,\ldots,k\}$, $T_i$ is the disjoint union of vertices and at most one edge.

\item[$(iii)$] $G[N_3]$ is bipartite.

\item[$(iv)$] $S_3 \subseteq I$, i.e., $S_3$ is an independent vertex set.

\item[$(v)$] If a vertex $t_i \in T_i$ sees two vertices in $T_j$, $i \neq j$, $i,j \in \{1,\ldots,k\}$, then $u_it_i \in M$ is an $xy$-forced $M$-edge.

\end{enumerate}
\end{lemma}

{\bf Proof.} $(i)$: Holds by definition of $T_i$ and by the distance condition of a d.i.m.\ $M$.

$(ii)$: Holds by Observation \ref{obse:neighborhood}.

$(iii)$: Follows by Observation \ref{dimC3C5C7C4} $(i)$ since every odd cycle in $G$ must contain at least one $M$-edge, and by (\ref{noMedgesN3N4}).

$(iv)$: If $v \in S_3:= N_3 \setminus T_{one}$, i.e., $v$ sees at least two $M$-vertices then clearly, $v \in I$, and thus, $S_3 \subseteq I$ is an independent vertex set (recall that $I$ is an independent vertex set).

$(v)$: Suppose that $t_1 \in T_1$ sees $a$ and $b$ in $T_2$. Then, if $ab \in E$, $u_2,a,b,t_1$ induce a diamond in $G$. Thus, $ab \notin E$ and now,
$u_2,a,b,t_1$ induce a $C_4$ in $G$; the only possible $M$-edge for dominating $t_1a,t_1b$ is $u_1t_1$, i.e., $t_1=u'_1$.
\qed

\medskip

Thus, by $(v)$, from now on, we can assume that for every $i,j \in \{1,\ldots,k\}$, $i \neq j$, any vertex $t_i \in T_i$ sees at most one vertex in $T_j$.

\medskip

Then let us split the problem of checking if a d.i.m.\ $M$ with $xy$ exists into two cases:
The case $N_4 = \emptyset$ and the case $N_4 \neq \emptyset$.

\section{The case $N_4 = \emptyset$}\label{section:N4empty}

Throughout this section, we assume that $N_4 = \emptyset$.

\begin{lemma}\label{lemm:N4N5empty}
The following statements hold:
\begin{enumerate}
\item[$(i)$] For every edge $vw \in E$, $v,w \in N_3$, with $vu_i \in E$ and $wu_j \in E$, $|\{v,w\} \cap \{u'_i,u'_j\}| = 1$.  
\item[$(ii)$] For every edge $st \in E$ with $s \in S_3$ and $t \in T_i$, $t=u'_i$ holds, and thus $u_it$ is an $xy$-forced $M$-edge.
\end{enumerate}
\end{lemma}

{\bf Proof.} $(i)$: Since $N_4 = \emptyset$ and $vw \notin M$ (by (\ref{noMedgesN3N4}), $N_3$ does not contain any $M$-edge), $vw$ has to be dominated by exactly one of the $M$-edges $u_iu'_i$, $u_ju'_j$.

\medskip

$(ii)$: By Lemma \ref{lemm:structure2}, $S_3 \subseteq I$ and thus, by $(i)$, for the edge $st$ with $s \in S_3$, $t=u'_i$ holds.
\qed

\medskip

From now on, we can assume that $S_3$ is isolated in $N_3$. This means that every edge between $N_2$ and $N_3$ containing a vertex of $S_3$ is dominated; thus, we can assume that $S_3=\emptyset$. This means that for every $t \in N_3$, there is exactly one $i \in \{1,\ldots,k\}$ such that $u_it \in E$. Recall that $N_2=S_2=\{u_1,\ldots,u_k\}$.

\medskip

Let us observe that to check if a vertex set $W \subseteq T_{one}$ may be such that $W \subset V(M)$ (i.e.,
formed by the $M$-mates of some vertices of $S_2$) and to check the implications of this choice can be done by repeatedly applying forcing rules; the details are given in the following procedure which is correct by the above and which can be executed in polynomial time.\\

{\bf Procedure Extend[$W$-in-$M$]}

{\bf Given}: A vertex set $W \subseteq T_{one}$ and the vertex set $W' \subseteq S_2 \cup T_{one}$ formed by the vertices of those connected components of $G[S_2 \cup T_{one}]$ containing $W$. 

{\bf Task}: Return a proof that $G$ has no d.i.m.\ $M$ with $W \subset V(M)$, or return a partition of $T_{one} \cap W'$ into the set $T_{one,Col}$ of
{\em colored} vertices (by black or white) and the set $T_{one,Uncol}$ of {\em uncolored} vertices such that:  
\begin{enumerate}
\item[$(i)$] $T_{one,Col} \cojoin T_{one,Uncol}$ 
\item[$(ii)$] the set of black vertices of $T_{one,Col}$ and the set $S_{2,Col}$ of their respective neighbors in $S_2$ induce a d.i.m.\ of $G[S_{2,Col} \cup T_{one,Col}]$, and
\item[$(iii)$] the set of white vertices of $T_{one,Col}$ is that of vertices of $G[S_{2,Col} \cup T_{one,Col}]$ which are not in such a d.i.m.
\end{enumerate}

{\em Comment:} Once assumed that $W \subset V(M)$, the procedure colors vertices of $T_{one} \cap W'$ which should be in $V(M)$ black, and vertices of $T_{one} \cap W'$ which should be in $I$ white.

\medskip
\begin{itemize}
\item[ ] {\bf Step 1.} Color all vertices of $W$ black.

\item[ ] {\bf Step 2.} Color some vertices of $T_{one} \cap W'$ either black or white by repeatedly applying the following forcing rules:
\begin{itemize}
\item[(a)] set $X: = \emptyset$;

\item[(b)] Repeat
\begin{itemize}
\item[(b.1)] take a colored vertex of $(T_{one} \cap W') \setminus X$, say $v \in T_i \cap W'$, and set $X:= X \cup \{v\}$; 
\item[(b.2)] if $v$ is black, then color all neighbors of $v$ in $T_{one} \cap W'$ white, and color all vertices of $T_i \setminus \{v\}$ white;
\item[(b.3)] if $v$ is white, then color all neighbors of $v$ in $T_{one} \cap W'$ black.
\end{itemize}
until there is no colored vertex in $(T_{one} \cap W') \setminus X$.
\end{itemize}

\item[ ] {\bf Step 3.}  If referring to Step 2, a vertex of $T_{one} \cap W'$ should change its color, i.e., it is colored white (black, respectively) while being black (white, respectively), then return a proof that $G$ has no d.i.m.\ $M$ with $t_1 \in V(M)$. Otherwise, return a partition of $T_{one} \cap W'$ according to the Task (of the procedure). 
\end{itemize}

Let us say that Procedure Extend[$W$-in-$M$] is {\em complete} if it either returns a proof that $G$ has no d.i.m.\ $M$ with $W \subset V(M)$, or returns $T_{one,Uncol} = \emptyset$, and is {\em incomplete} otherwise. Note that Procedure Extend[$W$-in-$M$] may be incomplete. Furthermore note that a white vertex of $T_{one,Col}$ may have a neighbor in $S_2 \setminus S_{2,Col}$.

\medskip

Then let us focus on $G[S_2 \cup T_{one}]$. Only two cases are possible according to the following subsections \ref{subsec:N3noedges} and \ref{subsec:N3edges}:

\begin{itemize}
\item[4.1] $T_i$ \cojoin $T_j$
\item[4.2] $T_i$ sees $T_j$ for some $1 \leq i < j \leq k$
\end{itemize}

\subsection{There is no edge between $T_i$ and $T_j$ for $1 \leq i < j \leq k$}\label{noedgesN3}\label{subsec:N3noedges}

In this case the problem of checking if $M$ exists can be solved in polynomial time as follows:

For each vertex $t_i \in T_i$, for $i = 1,\ldots,k$, run Procedure Extend[$W$-in-$M$] with $W = \{t_i\}$ and choose a minimum finite weight solution (if such a solution exists) over $t \in T_i$. Note that Procedure Extend[$W$-in-$M$] with $W = \{t_i\}$ is complete (that can be easily checked since the connected component of $G[S_2 \cup T_{one}]$ containing $t_i$ is $G[\{u_i\} \cup T_i]$). Finally either return that $G$ has no d.i.m.\ $M$ with $xy$ or return $M$.

\subsection{There is an edge between $T_i$ and $T_j$ for some $1 \leq i < j \leq k$}\label{edgesN3}\label{subsec:N3edges}

Assume that there is an edge $t_it_j \in E$ between $t_i \in T_i$ and $t_j \in T_j$, for some $i,j \in \{1,\ldots,k\}$, $i \neq j$; without loss of generality, let $i=1$ and $j=2$ and $t_1t_2 \in E$. Let $G'$ be the subgraph of $G$ induced by the non-neighborhood of $t_1,t_2$.

\begin{lemma}\label{lemm:structure3}
The following statements hold for every $i \in \{3,\ldots,k\}$ in $G'$:
\begin{itemize}
\item[$(i)$] Each edge $e_i$ in $T_i$ misses each vertex in $\{T_{3}, \ldots, T_k\} \setminus \{T_i\}$.
\item[$(ii)$] Each vertex $t_i \in T_i$ sees at most one vertex in $\{T_{3}, \ldots, T_k\} \setminus \{T_i\}$.
\end{itemize}
\end{lemma}

{\bf Proof.}
$(i)$: Without loss of generality, suppose to the contrary that for an edge $t_it'_i \in E$ with $t_i,t'_i \in T_i$, there is a vertex $t_j \in T_j$ with $t_it_j \in E$. Then by Lemma \ref{lemm:structure2} $(iii)$, $t'_it_j \notin E$ but now, the subgraph of $G$ induced by $t_2,t_1,u_1,N_1,x,y,u_j,t_j,t_i,t'_i$ contains a $P_8$.

$(ii)$: By Lemma \ref{lemm:structure2} $(v)$, we can assume that no vertex in $T_i$ sees two vertices in $T_j$. Without loss of generality, suppose to the contrary that there is a vertex $t_i \in T_i$ which sees $t_j \in T_j$ and $t_q \in T_q$, $j \neq q$. Then again by Lemma \ref{lemm:structure2} $(iii)$, $t_jt_q \notin E$ but now, the subgraph of $G$ induced by $t_2,t_1,u_1,N_1,x,y,u_q,t_q,t_i,t_j$ contains a $P_8$.
\qed

\medskip

Let $Z$ be the graph with nodes $\{z_{3},\ldots,z_k\}$, where $z_i$ corresponds to $T_i$ for $i \in \{3$,$\ldots,k\}$, such that for $i \neq j$, $z_iz_j$ is an edge in $Z$ if and only if $T_i$ sees $T_j$ in $G$. Let us say that:
\begin{itemize}
\item[$(i)$] $T_i$ forms a {\em singleton-type} in $G[H]$ if the node of $Z$ corresponding to $T_i$ is an isolated node of $Z$.

\item[$(ii)$] $T_i$ and $T_j$ form an {\em edge-type} in $G[H]$ if $z_iz_j$ is an isolated edge of $Z$.

\item[$(iii)$] $T_i,T_{j_1},\ldots,T_{j_h}$ form a {\em star-type} in $G[H]$ if the nodes of $Z$ corresponding to $T_i,T_{j_1},\ldots,T_{j_h}$ form an isolated non-trivial star of $Z$ with center $T_i$, for $i,j_1,\ldots,j_h \in \{3,\ldots,k\}$. Let
\begin{itemize}
\item[ ] $T'_i := \{t_i \in T_i: t_i$ sees an element of $\{T_{j_1},\ldots,T_{j_h}\}\}$ and
\item[ ] $T'_{i,j} := \{t_i \in T_i: t_i$ sees an element of $T_j\}$ for $j \in \{j_1,\ldots,j_h\}$.
\end{itemize}
\end{itemize}

\begin{lemma}\label{lemm:structureZ}
Each component of $Z$ in $G'$ is either a singleton or an edge or a non-trivial star.
\end{lemma}

{\bf Proof.}
If for all $i \in \{3,\ldots,k\}$, $T_i$ sees at most one element of $\{T_{3}, \ldots, T_k\} \setminus \{T_i\}$, then the components of $Z$ are either singletons or edges, and Lemma \ref{lemm:structureZ} follows. Thus assume that there is an $i \in \{3,\ldots,k\}$ such that $T_i$ sees more than one element of $\{T_{3}, \ldots, T_k\} \setminus \{T_i\}$, say $T_i$ sees $T_{j_1},\ldots,T_{j_h}$, for some $\{j_1,\ldots,j_h\} \subseteq \{3,\ldots,k\} \setminus \{i\}$ with $h \ge 2$. Let us prove that the nodes of $Z$ corresponding to $T_i,T_{j_1},\ldots,T_{j_h}$ induce in $Z$ an isolated non-trivial star with center $T_i$; that will imply Lemma \ref{lemm:structureZ}.

Let $T'_i$ and $T'_{i,j}$ be as defined in $(iii)$ above. Then $T'_i = T'_{i,j_1} \cup \ldots \cup T'_{i,j_h}$ is a partition of $T'_i$ by Lemma \ref{lemm:structure3} $(ii)$. Moreover $T'_i$ misses $T_i \setminus T'_i$ by Lemma \ref{lemm:structure3} $(i)$.

\medskip

{\em Notation:} For a clear reading let us write $j_1 = \xi$ and $j_2 = \eta$.

\begin{clai}\label{claim1}
$T'_i \subset I$.
\end{clai}

{\em Proof.} By contradiction assume that a vertex from $T'_i$ is in $V(M)$, say a vertex $t_{i,\xi} \in T'_{i,\xi}$ without loss of generality, i.e., $t_{i,\xi}$ is the $M$-mate of $u_i$. Then $T'_{i,j} \subset I$ for all $j \in \{j_2,\ldots,j_h\}$ by Lemma \ref{lemm:structure2} $(i)$.
By definition of $T'_{i,\xi}$, $t_{i,\xi}$ sees a vertex $t'_{\xi} \in T_{\xi}$. Then, since $t_{i,\xi} \in V(M)$, we have $t'_{\xi} \in I$. Then by Lemma \ref{lemm:structure2} $(i)$ there is a vertex $t_{\xi} \in T_{\xi}$ such that $t_{\xi} \in V(M)$, namely the $M$-mate $u'_{\xi}$ of $u_{\xi}$:
In particular by Lemma~\ref{lemm:structure3} $(i)$ we derive that $t'_{\xi}$ misses $t_{\xi}$.

On the other hand by definition of $T'_{i,\eta}$, a vertex $t_{i,\eta} \in T'_{i,\eta}$ sees a vertex $t'_{\eta} \in T_{\eta}$.
Then since $t_{i,\eta} \in I$, one has $t'_{\eta} \in V(M)$, i.e., $t'_{\eta}$ is the $M$-mate $u'_{\eta}$ of $u_{\eta}$:
In particular by Lemma~\ref{lemm:structure3} $(i)$ we derive that $t_{i,\eta}$ misses $t_{i,\xi}$ but then, by Lemma \ref{lemm:structure3} $(ii)$ and by the above, $u_{\eta},t'_{\eta},t_{i,\eta},u_i,t_{i,\xi},t'_{\xi},u_{\xi},t_{\xi}$ induce a $P_8$.
This shows Claim \ref{claim1}.
\hfill $\diamond$

\medskip

Claim \ref{claim1} implies: $T_i \setminus T'_i \neq \emptyset$ and contains the $M$-mate of $u_i$ by Lemma \ref{lemm:structure2} $(i)$;
each vertex of $T'_{i,j}$, for $j \in \{j_1,\ldots,j_h\}$, sees exactly one vertex of $T_j$, namely the $M$-mate $u'_j$ of $u_j$ (in particular all vertices of $T'_{i,j}$ have the same neighborhood in $T_j$).

\begin{clai}\label{claim2}
The elements of $\{T_{j_1},\ldots,T_{j_h}\}$ miss each other.
\end{clai}

{\em Proof.} Without loss of generality, by symmetry let us only show that $T'_{\xi}$ misses $T'_{\eta}$.
By contradiction assume that there is an edge $t'_{\xi}t'_{\eta}$ between $T'_{\xi}$ and $T'_{\eta}$.
Let $t_{i,\eta} \in T'_{i,\eta}$ and let $t_{\eta} \in T'_{\eta}$ be the $M$-mate of $u_{\eta}$.
Then $t_{i,\eta}$ sees $t_{\eta}$ (by the above) and consequently: $t_{\eta} \neq t'_{\eta}$ by Lemma \ref{lemm:structure3} $(ii)$,
any $t_{i,\xi} \in T'_{i,\xi}$ misses $t'_{\eta}$ since they are both in $I$, $t_{\eta}$ misses $t'_{\eta}$ by Lemma \ref{lemm:structure3} $(i)$, and finally $t_{i,\xi}$ and $t_{\eta}$ miss $t'_{\xi}$ by Lemma \ref{lemm:structure3} $(ii)$. Then $u_{\xi},t'_{\xi},t'_{\eta},u_{\eta},t_{\eta},t_{i,\eta},u_i$ and any vertex of $T_i \setminus T'_i$ induce a $P_8$.
This completes the proof of Claim \ref{claim2}.
\hfill $\diamond$

\begin{clai}\label{claim3}
No element of $\{T_i,T_{j_1},\ldots,T_{j_h}\}$ sees any element of $\{T_{3}, \ldots, T_k\} \setminus \{T_i,T_{j_1},\ldots,T_{j_h}\}$.
\end{clai}

{\em Proof.} The fact holds true for $T_i$ by construction.
Without loss of generality by symmetry we only need to show that $T_{\eta}$ misses $T_{\zeta}$, where $\zeta \in \{3,\ldots,k\} \setminus \{i,j_1,\ldots,j_h\}$. Suppose to the contrary that there is an edge $t'_{\eta}t'_{\zeta}$ between $T_{\eta}$ and $T_{\zeta}$. Let $t_{i,\eta} \in T_{i,\eta}$ and let $t_{\eta} \in T_{\eta}$ be the $M$-mate of $u_{\eta}$. Then $t_{i,\eta}$ sees $t_{\eta}$ (by the above) and consequently: $t_{\eta} \neq t'_{\eta}$ by Lemma \ref{lemm:structure3} $(ii)$, $t_{i,\eta}$ misses $t'_{\eta}$ since they are both in $I$, $t_{\eta}$ misses $t'_{\eta}$ by Lemma \ref{lemm:structure3} $(i)$ , and finally $t_{i,\eta}$ and $t_{\eta}$ miss $t'_{\zeta}$ by Lemma \ref{lemm:structure3} $(ii)$. Then $u_{\zeta},t'_{\zeta},t'_{\eta},u_{\eta},t_{\eta},t_{i,\eta},u_i$ and any vertex of $T_i \setminus T'_i$ induce a $P_8$.
This completes the proof of Claim \ref{claim3}.
\hfill $\diamond$

\medskip

Now Claims \ref{claim1}, \ref{claim2}, and \ref{claim3} imply that the nodes of $Z$ corresponding to $T_i,T_{j_1},\ldots,T_{j_h}$ induce an isolated non-trivial star in $Z$. Thus Lemma \ref{lemm:structureZ} follows.
\qed

\medskip

According to Lemma \ref{lemm:structureZ}, let us focus on a connected component of $G[\{u_3,\ldots,u_k\} \cup T_3 \cup \ldots \cup T_k]$, say $Q = G[\{u_i,u_{j_1},\ldots,u_{j_h}\} \cup T_i \cup T_{j_1} \cup \ldots \cup T_{j_h}$], with $T_i,T_{j_1},\ldots,T_{j_h}$ inducing a (trivial or non-trivial) star in $Z$ with center $T_i$ (recall that the cardinality of the family $\{T_{j_1}, \ldots, T_{j_h}\}$ may be even equal to 0 or to 1).

\medskip

Then let us observe that, to compute a minimum weight d.i.m.\ of $Q$ %$G[\{u_i,u_{j_1},\ldots,u_{j_h}\} \cup T_i \cup T_{j_1} \cup \ldots \cup T_{j_h}$]
(if it exists), say $M'$, with $\{u_i,u_{j_1},\ldots,u_{j_h}\} \in V(M')$, and with a fixed vertex $t_i \in T_i$ being in $V(M')$ (i.e., being the $M$-mate of $u_i$), can be done by the following procedure which is correct by the above and which can be executed in polynomial time.

\medskip

{\bf Step 1.} Run Procedure Extend[$W$-in-$M$] with $W = \{t_i\}$.

{\bf Step 2.} If it returns $T_{one,Uncol} = \emptyset$ (i.e., if it is complete) then we are done.

{\bf Step 3.} If it is incomplete and returns a partition of $T_{one} \cap W'$, 
namely $\{T_{one,Col},T_{one,Uncol}\}$, with $T_{one,Uncol} \neq \emptyset$ then we can easily color the vertices of $T_{one,Uncol}$ such that black vertices are finally the $M$-mates of $\{u_i,u_{j_1},\ldots,u_{j_h}\}$: in fact by construction and by the above, we have $T_{one,Uncol} \subseteq T_{j_1} \cup \ldots \cup T_{j_h}$, and in particular, for each $j \in \{j_1,\ldots,j_h\}$, $T_{one,Uncol} \cap T_j$  has a co-join to $T_{one,Col} \cup (T_{one,Uncol} \setminus T_j)$ and induces a subgraph with at most one isolated edge $e_j = ab$ (say with $w(au_j) \leq w(bu_j)$) and isolated vertices; now, if $ab$ exists then we color vertex $a$ black, and if $ab$ does not exist then we color exactly one vertex $t_j \in T_{one,Uncol} \cap T_j$ black such that $w(t_ju_j) \leq w(tu_j)$ for $t \in T_{one,Uncol} \cap T_j$.

\medskip

Then let us summarize the above (recall that without loss of generality, there is an edge between $T_1$ and $T_2$). In this case the problem of checking if a d.i.m.\ $M$ exists can be solved in polynomial time by Lemma \ref{lemm:structureZ} as follows:

\begin{itemize}
\item[(a)] For a vertex $t_1 \in T_1$ such that $t_1$ has a neighbor $t_2 \in T_2$, and for each vertex $t'_2 \in T_2$ such that $t'_2$ is a non-neighbor of $t_1$ in $T_2$ (such a non-neighbor may not exist), do as follows:

\begin{itemize}
\item[(a.1)] Run Procedure Extend[$W$-in-$M$] with $W = \{t_1,t'_2\}$. If it returns a partition of $T_{one} \cap W'$, namely $\{T_{one,Col},T_{one,Uncol}\}$, then go to Step (a.2). Note that $T_{one,Uncol} \subseteq T_3 \cup \ldots \cup T_k$, and that more generally $G[(S_2 \setminus S_{2,Col}) \cup T_{one,Uncol}]$ is a subgraph of $G[\{u_3,\ldots,u_k\} \cup T_3 \cup \ldots \cup T_k \setminus (N(t_1) \cup N(t_2))]$.

\item[(a.2)] For each connected component $Q$ of $G[(S_2 \setminus S_{2,Col}) \cup T_{one,Uncol}]$ do as follows: for each $q \in Q$, compute a minimum finite weight d.i.m. of $Q$ (if it exists), say $M'$, with $\{u_i,u_{j_1},\ldots,u_{j_h}\} \in V(M)$, and with $q$ being in $V(M')$, as shown above, and choose a minimum weight solution (if a solution exists) over $q \in Q$.

\item[(a.3)] Obtain a minimum finite weight d.i.m.\ containing $t_1$ and $t'_2$ by collecting those solutions found in steps (a.1)-(a.2) (if those solutions exist).
\end{itemize}

\item[(b)] Analogously, for a vertex $t_2 \in T_2$ such that $t_2$ has a neighbor $t_1 \in T_1$, and for each $t'_1 \in T_1$ such that $t'_1$ is a non-neighbor of $t_2$ in $T_1$ (such a non-neighbor may not exist), proceed as in steps (a.1), (a.2), (a.3), by symmetry.

\item[(c)] Choose a minimum finite weight solution (if such a solution exists) among those found in steps (a)-(b) respectively for $(t_1,t'_2) \in T_1 \times T_2$ and for $(t'_1,t_2) \in T_1 \times T_2$ as defined above and return $M$, or return that $G$ has no d.i.m.\ $M$ with $xy$.
\end{itemize}

\section{The case $N_4 \neq \emptyset$}\label{section:nonempty}

The aim of this section is to reduce the graph step by step so that finally $N_4 = \emptyset$.

\subsection{Components of $N_4$}\label{subsec:N4components}

The aim of this subsection is to reduce the graph so that $N_4$ becomes an independent set. For showing this, we need several lemmas:

\begin{lemma}\label{lemm:structure4.1}
$N_4$ is $P_3$-free.
\end{lemma}

{\bf Proof.} Suppose to the contrary that there is a $P_3$ in $G$ with vertices $a,b,c \in N_4$ and edges $ab$ and $bc$. Let $a'$ be a neighbor of $a$ in $N_3$. 
This proof follows the principle of the proof of (\ref{N6empty}). Let us recall that $\{r,x,y\}$ induces a $P_3$ with edge $rx$. Then, to avoid a $P_8$ in the subgraph induced by $c,b,a,a',N_2 \cup N_1,x,y$ (in detail, denoted as $a''$ a neighbor of $a' \in N_2$, and denoted as $r''$ a neighbor of $a''$ in $N_1$, the $P_8$ would be induced by $c,b,a,a',a''$, and: either $r'',x,y$ if $r'' = r$, or $r'',x,r$ if $r'' \neq r$), $a'$ sees either $b$ or $c$ but not both since $G$ is diamond-free.

\medskip

{\bf Case 1.} $a'$ sees $c$ (and misses $b$).

\medskip

Then $a',a,b,c$ induce a $C_4$ in $G$, and thus, by Observation \ref{dimC3C5C7C4} $(ii)$, either $a',b \in V(M)$ (and $a,c \in I$), or $a,c \in V(M)$ (and $a',b \in I$).

Assume first that $a',b \in V(M)$ (and $a,c \in I$). Let $b^*$ be the $M$-mate of $b$. Since by (\ref{noMedgesN3N4}), no edge between $N_3$ and $N_4$ is in $M$, it follows that $b^* \in N_4 \cup N_5$ but then to avoid a $P_8$ (in the subgraph induced by $b^*,b,a,a'$, $N_2 \cup N_1,x,y$), $a$ sees $b^*$, and to avoid a $P_8$ (in the subgraph induced by $b^*,b,c,a'$,$N_2 \cup N_1,x,y$), $c$ sees $b^*$ but now $a,b,b^*,c$ induce a diamond which is a contradiction.

Thus, assume that $a,c \in V(M)$ (and $a',b \in I$). Let $a^*,c^*$ respectively be the $M$-mates of $a$ and $c$. Since by (\ref{noMedgesN3N4}), no edge between $N_3$ and $N_4$ is in $M$, it follows that $a^*,c^* \in N_4 \cup N_5$. Let $b'$ be a neighbor of $b$ in $N_3$; clearly, $b' \neq a'$. Then $b' \in V(M)$ (since $b \in I$). Then $b'$ misses $c,c^*$, and thus a $P_8$ arises (in the subgraph induced by $c^*,c,b,b'$, $N_2 \cup N_1,x,y$ if $bc^* \notin E$ or in the subgraph induced by $a^*,a,b,b'$, $N_2 \cup N_1,x,y$ if $bc^* \in E$; in that case, $ba^* \notin E$ since $G$ is butterfly-free). Thus, Case 1 is impossible.

\medskip

{\bf Case 2.} $a'$ sees $b$ (and misses $c$).

\medskip

Let $c'$ be a neighbor of $c$ in $N_3$. By symmetry with respect to Case 1, $c'$ sees $b$ (and misses $a$). Then the subgraph induced by $a',a,b,c,c'$ contains a butterfly or a diamond. Thus, also Case 2 is impossible which completes the proof of Lemma \ref{lemm:structure4.1}.
\qed

\medskip

Recall that a graph is $P_3$-free if and only if it is the disjoint union of complete graphs. Since we can assume that $G$ is $K_4$-free, we have:
\begin{corollary}\label{N4compon}
The components of $N_4$ are triangles, edges or isolated vertices.
\end{corollary}

\subsubsection{Triangles in $N_4$}

\begin{lemma}\label{lemm:N4triangle}
Let $H$ be a triangle component of $N_4$ with vertices $a,b,c$, edges $ab,ac,bc$,
and let $A := N(a) \cap N_3$, $B := N(b) \cap N_3$, and $C := N(c) \cap N_3$. Then the following statements hold:

\begin{itemize}
\item[$(i)$] $A,B,C$ are pairwise disjoint independent sets.

\item[$(ii)$] $H \cojoin N_5$.

\item[$(iii)$] $(A \cup B \cup C) \cap S_3 = \emptyset$.

\item[$(iv)$] There exists $j, 1 \le j \le k$, such that $A \cup B \cup C \subseteq T_j$.
\end{itemize}
\end{lemma}

{\bf Proof.}
$(i)$: Holds by Observation \ref{obse:neighborhood} since $G$ is ($K_4$, diamond, butterfly)-free.

\medskip

$(ii)$: Without loss of generality, suppose to the contrary that there is a neighbor of $c$ in $N_5$, say $z$. Then $z$ misses $b$, otherwise a diamond or a $K_4$ arises. Let $b'$ be a neighbor of $b$ in $N_3$. Then by $(i)$, $b'$ misses $c$ but now, a $P_8$ arises (with $z,c,b,b'$, $N_2 \cup N_1$ and a $P_3$ containing $x,y$).

\medskip

$(iii)$: Without loss of generality, suppose to the contrary that there is a vertex $a' \in A \cap S_3$, say $a'u_1 \in E$ and $a'u_2 \in E$. Let $b' \in B$ and $c' \in C$. If $b' \in S_3$ and $c' \in S_3$ as well, then $a,b,c \in V(M)$ (recall that by Lemma \ref{lemm:structure2} $(iv)$, $S_3 \subseteq I$). Thus, assume that $b' \notin S_3$, i.e., $b'$ has only one neighbor in $u_1,\ldots,u_k$ and thus, $b'$ misses $u_1$ or $u_2$, say $b'u_1 \notin E$. Then if $a'b' \notin E$, the subgraph induced by $b',b,a,a',u_1,N_1,x,y$ contains a $P_8$, and if $a'b' \in E$, the subgraph induced by $c,b,b',a',u_1,N_1,x,y$ contains a $P_8$ which is a contradiction.

\medskip

$(iv)$: The proof is similar to that of $(iii)$; without loss of generality, let $a' \in A$ see $u_1$ and suppose to the contrary that there is a vertex $b' \in B$ missing $u_1$. Then
if $a'b' \notin E$, the subgraph induced by $b',b,a,a',u_1,N_1,x,y$ contains a $P_8$, and if $a'b' \in E$, the subgraph induced by $c,b,b',a',u_1,N_1,x,y$ contains a $P_8$ which is a contradiction.
\qed

\medskip

As in Lemma \ref{lemm:N4triangle}, for a triangle $a_ib_ic_i$ in $N_4$ let $A_i$ ($B_i, C_i$, respectively) denote the neighborhood of $a_i$ (of $b_i, c_i$, respectively) in $N_3$.

\begin{corollary}\label{commonindextriangles}
There exists $j, 1 \le j \le k$, such that for all triangles $a_ib_ic_i$ in $N_4$, $A_i \cup B_i \cup C_i \subseteq T_j$.
\end{corollary}

{\bf Proof.} Let $a_1b_1c_1$ and $a_2b_2c_2$ be two triangles in $N_4$ such that, without loss of generality, $A_1 \cup B_1 \cup C_1 \subseteq T_1$.
If there is a vertex in $A_2 \cup B_2 \cup C_2 \setminus T_1$, say $a'_2 \in A_2$ with $a'_2u_1 \notin E$ then by Lemma~\ref{lemm:N4triangle}, a $P_8$ arises.
Thus, $A_2 \cup B_2 \cup C_2 \subseteq T_1$ holds as well.
\qed

\medskip

From now on, without loss of generality, suppose that for every triangle $a_ib_ic_i$ in $N_4$, $A_i \cup B_i \cup C_i \subseteq T_1$. Assume that for the triangle $a_1b_1c_1$, the $M$-edge is $b_1c_1 \in M$. Then $A_1=\{u'_1\}$ since otherwise, if there is $a' \in A_1$ with $a' \neq u'_1$ then the edge $aa' \in E$ is not dominated by $M$.
Since every triangle contains exactly one $M$-edge, this implies that one of the sets $A_2,B_2,C_2$ is equal to $\{u'_1\}$, say $A_2=\{u'_1\}$ which forces the $M$-edge $b_2c_2 \in M$ and similarly for every triangle $a_ib_ic_i$ in $N_4$.

\medskip

Thus, if there is a triangle in $N_4$, we have to consider three possible cases according to the $M$-edges in the triangles (which in each of the cases can be considered as $xy$-forced).

\subsubsection{Edges in triangle-free $N_4$}

From now on, we can assume that $N_4$ is triangle-free. If component $H$ in $N_4$ is not a triangle then by Lemma \ref{lemm:structure4.1}, $H$ is either a vertex or an edge.

\begin{lemma}\label{lemm:structure4.3}
Let $H$ be a component of $N_4$ and assume that $H \cojoin N_5$. Then we have:
\begin{itemize}
\item[$(i)$] If $H=\{h\}$ then $h \in I$.
\item[$(ii)$] If $H=\{a,b\}$ with $ab \in E$ then $ab \in M$ and thus, $ab$ is an $xy$-forced $M$-edge.
\end{itemize}
\end{lemma}

{\bf Proof.}
The lemma follows by (\ref{noMedgesN3N4}) - none of the edges in $N_3$ and between $N_3$ and $N_4$ is in $M$.
\qed

\medskip

From now on, we can assume that $N_4$ is triangle-free and every edge in $N_4$ has a neighbor in $N_5$.
If $uv$ is an edge in $N_4$ then by (\ref{edgeN4N3neighb}), we can assume that $u$ and $v$ do not have a common neighbor in $N_3$; let $u' \in N_3$ ($v' \in N_3$, respectively) be a neighbor of $u$ (of $v$, respectively).

\begin{lemma}\label{lemm:structure5.2}
Let edge $ab \in E$ be a component $H$ in $N_4$ $($i.e., $\{a,b\} \cojoin (N_4 \setminus \{a,b\}))$ and let $c \in N_5$ be a neighbor of $ab$.
Let $A := N(a) \cap N_3$ and $B := N(b) \cap N_3$. Then the following statements hold:
\begin{itemize}
\item[$(i)$] Any neighbor $c \in N_5$ of $ab$ must see both of $a$ and $b$.

\item[$(ii)$] $A \cap B = \emptyset$ and $A,B$ are independent sets.

\item[$(iii)$] For all $a' \in A$ and $b' \in B$, $N(a') \cap N_2 = N(b') \cap N_2$.

\item[$(iv)$] If there is $a' \in A$ with $|N(a') \cap N_2| \ge 2$ $($there is $b' \in B$ with $|N(b') \cap N_2| \ge 2$, respectively$)$,  then $A \cojoin B$ and $ab$ is an $xy$-forced $M$-edge.

\item[$(v)$]
Otherwise, if for all $a' \in A$, $|N(a') \cap N_2| =1$ and  for all $b' \in B$, $|N(b') \cap N_2| =1$
then there is an index $i, 1 \le i \le k$ such that $A \cup B \subseteq T_i$.

\end{itemize}
\end{lemma}

{\bf Proof.}
$(i)$: If a neighbor $c \in N_5$ of $ab$ sees only one of $a$ and $b$, say $bc \in E$ and $ac \notin E$, then there is a $P_8$ in the subgraph induced by $c,b,a,a'$, $N_2 \cup N_1$ and a $P_3$ containing $x,y$. Thus, we can assume that each edge component in $N_4$ is contained in such a triangle with a common neighbor in $N_5$.

\medskip

$(ii)$: By (\ref{edgeN4N3neighb}), we can assume that $a$ and $b$ do not have a common neighbor in $N_3$. Moreover, since $a$ and $b$ have the common neighbor $c \in N_5$, a common neighbor of $a$ and $b$ in $N_3$ would lead to a diamond. Thus, $A \cap B = \emptyset$. Moreover, $A$ and $B$ are independent sets since otherwise, there is a butterfly in $G$.

\medskip

$(iii)$: Without loss of generality, suppose to the contrary that $a' \in A$ sees $u_1$ and $b' \in B$ misses $u_1$. Then
if $a'b' \in E$, a $P_8$ arises in the subgraph induced by $c,b,b',a',u_1,N_1,x,y$, and if $a'b' \notin E$, a $P_8$ arises in the subgraph induced by $b',b,a,a',u_1,N_1,x,y$.

\medskip

$(iv)$: Without loss of generality, assume that $a' \in A$ sees $u_1$ and $u_2$. Then by $(iii)$ each vertex of $A \cup B$ sees $u_1$ and $u_2$. Then $A \cojoin B$, since otherwise a diamond arises. Moreover, since $\{u_1,a',u_2,b'\}$ induce a $C_4$, $a' \neq u'_1$ and $a' \neq u'_2$, and thus, for the $C_5$ induced by $\{u_1,a',b',a,b\}$ (with $b' \in B$), exactly one edge is in $M$ (recall Observation \ref{dimC3C5C7C4} $(i)$ for $C_5$). Then, since $a',b' \in I$ (as they are in $S_3$), the only possible way is that $ab \in M$.

\medskip

$(v)$: It follows by statement $(iii)$.
\qed

\medskip

According to Lemma \ref{lemm:structure5.2} $(iv)$-$(v)$, in what follows let us assume that, for any triangle $abc$ with an edge $ab$ in $N_4$ and $c \in N_5$, $A \cup B \subseteq T_j$ for some index $j$, $1 \leq j \leq k$.

\begin{lemma}\label{lemm:2trianglesN4N5}
Let $a_1b_1$ and $a_2b_2$ be distinct edge components in $N_4$ such that $a_1b_1c_1$ and $a_2b_2c_2$ are triangles with $c_1,c_2 \in N_5$, and denote by $A_i$ $(B_i$, respectively$)$ the neighborhood of $a_i$ $(b_i$, respectively$)$, $i=1,2$, in $N_3$.
Then there is an index $j, 1 \le j \le k$ such that $A_1 \cup B_1 \cup A_2 \cup B_2 \subseteq T_j$.
\end{lemma}

{\bf Proof.} Clearly, $c_1 \neq c_2$ since otherwise there is a butterfly in $G$. Now, if there are two such triangles, say $a_1b_1c_1$ and $a_2b_2c_2$ such that without loss of generality, there are $a'_1 \in A_1$ with $u_1a'_1 \in E$ and $a'_2 \in A_2$ with $u_2a'_2 \in E$ then a $P_8$ arises.
\qed

\medskip

Let $\{a_1b_1c_1, \ldots, a_{\ell}b_{\ell}c_{\ell}\}$, $\ell \le m$, be the set of all triangles with an edge $a_ib_i$ in $N_4$ and $c_i \in N_5$. As above, denote by $A_i$ $(B_i$, respectively$)$ the neighborhood of $a_i$ $(b_i$, respectively$)$, in $N_3$.

Without loss of generality, assume that $u_1$ is a common $N_2$-neighbor of $A_i$ and $B_i$, $i \in \{1,\ldots,{\ell}\}$. Now there are at most $n$ (where $n = |V|$) possible cases for $u_1u'_1 \in M$ and the $M$-edges in the triangles according to the property whether the $M$-mate $u'_1$ of $u_1$ is in $A_1 \cup B_1 \cup \ldots \cup A_{\ell} \cup B_{\ell}$ or not (which implies the other $M$-edges in the triangles):

\begin{corollary}\label{2N4-1N5-triangleMimplic}
\mbox{ }
\begin{itemize}
\item[$(i)$] If for $i \in \{1,\ldots,\ell\}$ and for $a'_i \in A_i$, $u_1a'_i \in M$, then:
\begin{itemize}
\item for all $j$ such that $a'_i \not \in A_j$ and $a'_i \not \in B_j$, it follows that $a_jb_j \in M$;
\item for all $j$ such that $a'_i \in A_j$ and $a'_i \not \in B_j$, it follows that $b_jc_j \in M$;
\item for all $j$ such that $a'_i \not \in A_j$ and $a'_i \in B_j$, it follows that $a_jc_j \in M$.
\end{itemize}

Likewise, by symmetry, if for $i \in \{1,\ldots,\ell\}$ and for $b'_i \in B_i$, $u_1b'_i \in M$, the corresponding implications follow.

\item[$(ii)$] If for all $i \in \{1,\ldots,{\ell}\}$ and for all $(a'_i,b'_i) \in A_i \times B_i$, neither $u_1a'_i \in M$ nor $u_1b'_i \in M$ then for all $i \in \{1,\ldots,{\ell}\}$, $a_ib_i \in M$.
\end{itemize}
\end{corollary}

Subsequently, we can assume that $N_4$ is an independent set.

\subsection{Components of $N_5$}\label{subsec:N5components}

Throughout this subsection, let $H$ be a component in $N_5$. Recall that we can assume that $N_4$ is an independent set.

\begin{lemma}\label{lemm:structure5}
The following statements hold:

\begin{enumerate}
\item[$(i)$] For every neighbor $u \in N_4$ of any vertex of $H$, $ u \join H$ holds.
\item[$(ii)$] $H$ is either a single vertex or an edge.
\end{enumerate}
\end{lemma}

{\bf Proof.} $(i)$: It follows since otherwise a $P_8$ arises (with a $P_3$ containing $x,y$).

$(ii)$: It follows by statement $(i)$ and since $G$ is (diamond, $K_4$)-free.
\qed

\medskip

Now we have two cases which will be examined in the following subsections.

\subsubsection{$H$ is an edge, say $h_1h_2$}

\begin{lemma}\label{lemm:structure5.1}
Let $h_1h_2 \in E$ be an edge in $N_5$, let $c \in N_4$ be a common neighbor of $h_1,h_2$ and let $N(c) \cap N_5$ contain another vertex $h \notin \{h_1,h_2\}$. Then:
\begin{enumerate}
\item[$(i)$] $N(c) \cap N_5$ is formed by the disjoint union of vertices and edge $h_1h_2 \in E$, and is isolated in $N_5$.
\item[$(ii)$] If without loss of generality, $w(h_1c)\le w(h_2c)$ then $h_1c \in M$ is an $xy$-forced $M$-edge.
\end{enumerate}
\end{lemma}

{\bf Proof.} $(i)$: By Observation \ref{obse:neighborhood}, $N(c) \cap N_5$ is formed by the disjoint union of vertices and at most one edge, namely $h_1h_2 \in E$.

For showing that $N(c) \cap N_5$ is isolated in $N_5$, suppose to the contrary that there is an edge between $N(c) \cap N_5$ and $N(d) \cap N_5$ for some $d \in N_4$, $d \neq c$. Then there are $h \in N(c) \cap N_5$ and $h' \in (N(d) \setminus N(c)) \cap N_5$ such that $hh' \in E$. Then, by Lemma \ref{lemm:structure5} $(i)$, $ch' \in E$ which is a contradiction. 

\medskip

$(ii)$: By Observation \ref{dimC3C5C7C4} $(i)$, any triangle contains exactly one $M$-edge. We claim that the $M$-edge in the triangle $h_1h_2c$ 
must be either $h_1c$ or $h_2c$: Suppose to the contrary that $h_1h_2 \in M$. Then in order to dominate the edge $hc$, we need another neighbor $c' \in N_4$ of $h$ such that $c'h \in M$ (clearly, $cc' \notin E$). Now for any neighbor $d \in N_3$ of $c$, $d$ sees $c'$, since otherwise a $P_8$ arises (with $c',h,c,d$, $N_2 \cup N_1$ and a $P_3$ containing $x,y$) but then $d,c,h,c'$ induce a $C_4$ with $hc' \in M$ which is a contradiction to Observation \ref{dimC3C5C7C4} $(ii)$. Thus, either $h_1c \in M$ or $h_2c \in M$ and by the weight condition we can assume that $h_1c \in M$ is an $xy$-forced $M$-edge. 
\qed

\medskip

From now on, we can assume that for every $v \in N_4$, $N(v) \cap N_5$ is either an edge or an independent set. Subsequently, we first consider the case when $N(v) \cap N_5$ is an edge.

\begin{lemma}\label{lemm:structure5.3.1}
The following statements hold:
\begin{itemize}
\item[$(i)$] $|N(H) \cap N_4| = 1$, say $N(H) \cap N_4 = \{c\}$.
\item[$(ii)$] $N(c) \cap N_3$ is an independent set.
\end{itemize}
\end{lemma}

{\bf Proof.} $(i)$: By Lemma \ref{lemm:structure5} $(i)$, $N(h_1) \cap N_4 = N(h_2) \cap N_4$. Let $c \in N(h_1) \cap N_4$.
If $h_1$ has another neighbor $c' \neq c$ in $N_4$ then by Lemma \ref{lemm:structure5} $(i)$ (and by the assumption that $N_4$ is an independent set), $c'h_2 \in E$, and thus $h_1,h_2,c,c'$ induce a diamond which is a contradiction.

\medskip

$(ii)$: It follows by Observation \ref{obse:neighborhood} since otherwise, there is a butterfly in $G$.
\qed

\medskip

Without loss of generality assume that $w(h_1c) \leq w(h_2c)$. Then let:
\begin{itemize}
\item[ ] $D := N(c) \cap N_3$ (then by Lemma \ref{lemm:structure5.3.1} $(ii)$, $D$ is an independent set);
\item[ ] $D_i := T_i \cap D$, for $i \in \{1,\ldots,k\}$.
\end{itemize}

\begin{lemma}\label{lemm:structure5.3.2}
If $D \cap S_3 \neq \emptyset$ or $|D_i| \geq 2$ for some $i \in \{1,\ldots,k\}$, then $h_1c \in M$ is an $xy$-forced $M$-edge.
\end{lemma}

{\em Proof.} First assume that $D \cap S_3 \neq \emptyset$: Since $S_3 \subseteq I$ by Lemma \ref{lemm:structure2} $(iv)$, it follows that $c \in V(M)$, and then since $h_1h_2c$ is a triangle the assertion follows.

If $|D_i| \geq 2$ for some $i \in \{1,\ldots,k\}$ then for every $d \in D_i$, the edges $u_id$ belong to a $C_4$; then, since $u_i \in V(M)$, by Observation \ref{dimC3C5C7C4} $(ii)$ it follows that $D_i \subseteq I$, and then $c \in V(M)$, and since $h_1h_2c$ is a triangle, Lemma \ref{lemm:structure5.3.2} has been shown.
\qed

\medskip

According to Lemma \ref{lemm:structure5.3.2}, in what follows let us assume that $D \cap S_3 = \emptyset$ (i.e., $D \subseteq T_{one}$), and that $|D_i| \leq 1$ for all $i \in \{1,\ldots,k\}$.

Let $\{a_1b_1c_1, \ldots, a_{\ell}b_{\ell}c_{\ell}\}$, be the set of all triangles with $a_i \in N_4$ and $b_i,c_i \in N_5$.  Without loss of generality, let $w(a_ib_i) \le w(a_ic_i)$.
Clearly, $a_i \neq a_j$ for $i \neq j$ since otherwise there is a butterfly in $G$, and $a_ia_j \notin E$ since we can assume that $N_4$ is an independent set.

Similarly as for triangles in $N_4$ and for triangles with an edge in $N_4$, we are going to show that there are only polynomially many possible cases for $M$-edges in these triangles. Clearly, either $a_ib_i \in M$ or $b_ic_i \in M$ since $a_ib_ic_i$ is a triangle, $b_ic_i$ is a component in $N_5$ having exactly one neighbor in $N_4$, namely $a_i$, and $w(a_ib_i) \le w(a_ic_i)$.

Let $d_i$ denote a neighbor of $a_i$ in $N_3$. By Lemma \ref{lemm:structure5.3.2}, we can assume that every $d_i$ sees only one of $u_1,\ldots,u_k$.

\begin{lemma}\label{lemm:2trianglesN4N5edgeN5C4}
Let $a_1b_1c_1$ and $a_2b_2c_2$ be triangles as above with $b_1,b_2,c_1,c_2 \in N_5$, and denote by $d_i$ a neighbor of $a_i$, $i=1,2$, in $N_3$.
If $d_1 \in T_1$ and $d_2 \in T_2$ then $d_1,d_2,a_1,a_2$ induce a $C_4$ in $G$.
\end{lemma}

{\bf Proof.}
First let us show that $d_1d_2 \not \in E$. Assume to the contrary that $d_1d_2 \in E$. Then $d_2$ misses $a_1$, since otherwise a butterfly arises. Let us recall that $\{r,x,y\}$ induces a $P_3$ with edge $rx$. Then there is a $P_8$ with $b_1,a_1,d_1,d_2,u_2$, $N_1$ and $x,y$ which is a contradiction. Thus $d_1d_2 \not \in E$. 

Since there is no $P_8$ in the subgraph induced by $b_1,a_1,d_1,u_1,N_1,u_2,d_2,a_2,b_2$, it follows that either $d_1a_2 \in E$ or $d_2a_1 \in E$. We claim that $d_1a_2 \in E$ if and only if $d_2a_1 \in E$: In fact, if $d_1a_2 \in E$ and $d_2a_1 \not \in E$, then a $P_8$ is induced by $b_1,a_1,d_1,a_2,d_2,u_2$, a vertex of $N_1$, and $x$ or $y$; the other implication can be shown similarly by symmetry. Then a $C_4$ is induced by $d_1,d_2,a_1,a_2$.
\qed

\medskip

Now the $C_4$ leads to the fact that $a_1b_1 \in M$ if and only if $a_2b_2 \in M$. We say that two triangles $a_1b_1c_1$ and $a_2b_2c_2$ are {\em $C_4$-connected} if there are $d_1,d_2$ as above such that $d_1,d_2,a_1,a_2$ induce a $C_4$ in $G$,
and we say that a set of such triangles is a {\em $C_4$-connected component} if there is a sequence of such $C_4$-connected pairs reaching all of them. Obviously, for such a component, there are only two possibilities for $M$-edges.

\medskip

Then let us focus on triangles which are not in such a $C_4$-connected component. Similarly as for Lemma \ref{lemm:2trianglesN4N5}, we claim:

\begin{lemma}\label{lemm:2trianglesN4N5edgeN5}
Let $a_1b_1c_1$ and $a_2b_2c_2$ be triangles as above with $b_1,b_2,c_1,c_2 \in N_5$, and denote by $d_i$ a neighbor of $a_i$, $i=1,2$, in $N_3$.
Assume that $a_1b_1c_1$ and $a_2b_2c_2$ are not $C_4$-connected. Then there is an index $j, 1 \le j \le k$ such that $d_1,d_2 \in T_j$.
\end{lemma}

{\bf Proof.}
If there are two such triangles $a_1b_1c_1$ and $a_2b_2c_2$ such that $d_1,d_2$ do not have a common neighbor in $N_2$, say without loss of generality, $u_1d_1 \in E$ and $u_2d_2 \in E$ but $u_1d_2 \notin E$ and $u_2d_1 \notin E$ then a $P_8$ arises.
\qed

\medskip

Let $\{a_1b_1c_1, \ldots, a_{\ell}b_{\ell}c_{\ell}\}$, be the set of all triangles, which are not in a $C_4$-connected component, with an edge $b_ic_i$ in $N_5$, and let $A_i$ be the neighborhood of $a_i$ in $N_3$. Assume without loss of generality that $w(a_ib_i) \le w(a_ic_i)$.
Without loss of generality, assume that $u_1$ is the only $N_2$-neighbor of $A_i$, $i \in \{1,\ldots,{\ell}\}$.
Now there are at most $n$ (where $n = |V|$) possible cases for $u_1u'_1 \in M$ and the $M$-edges in the triangles:

\begin{corollary}\label{1N4-2N5-triangleMimplic}
\mbox{ }
\begin{itemize}
\item[$(i)$] If for $i \in \{1,\ldots,l\}$ and for $d_i \in A_i$, $u_1d_i \in M$ then for all $j$ such that $d_i \in A_j$ it follows that $b_jc_j \in M$, and for all $j$ such that $d_j \not \in A_j$ it follows that $a_jb_j \in M$.
\item[$(ii)$] If for all $i \in \{1,\ldots,{\ell}\}$ and for all $d_i \in A_i$, $u_1d_i \notin M$ then for all $i \in \{1,\ldots,{\ell}\}$, $a_jb_j \in M$.
\end{itemize}
\end{corollary}

Subsequently, we can assume that $N_5$ is an independent set.

\subsubsection{$H$ is a single vertex, say $h$}

\begin{lemma}\label{lemm:structure5.0}
If $|N(h) \cap N_4| \geq 2$ then $h \in I$.
\end{lemma}

{\bf Proof.} Let us recall that $N(h) \cap N_4$ is an independent set. Let $a,b \in N(h) \cap N_4$, $a \neq b$, and let $c \in N_3$ be a neighbor of $a$. 
Then $bc \in E$ since otherwise a $P_8$ with $b,h,a,c$, $N_2 \cup N_1$ and $x,y$ arises. This holds for every pair of neighbors $a,b \in N(h) \cap N_4$ of $h$. Thus every edge incident to $h$ is in a $C_4$, i.e., $h \in I$.
\qed

\begin{lemma}\label{lemm:structure5.1edgeless}
Assume that $|N(h) \cap N_4| = 1$, say $N(h) \cap N_4=\{v_4\}$. Then $v_4v_5 \in M$ is an $xy$-forced $M$-edge for some $v_5 \in N(v_4) \cap N_5$ having exactly one neighbor in $N_4$, depending on the best alternative.
\end{lemma}

{\bf Proof.} Since we can assume now that $N_5$ is an independent set, since by (\ref{noMedgesN3N4}) no edge between $N_3$ and $N_4$ is in $M$, since by Lemma \ref{lemm:structure5.0}, $v_4u \not \in M$ for every $u \in N_5$ having more than one neighbor in $N_4$, and since $v_4$ is the only neighbor of $h$ in $N_4$, it follows that $v_4v_5 \in M$ for some $v_5 \in N(v_4) \cap N_5$  having exactly one neighbor in $N_4$ (depending on the best alternative; possibly $h=v_5$) since otherwise, the edge $v_4h$ is not dominated.
\qed

\medskip

Thus, from now on, we can assume that every vertex of $N_5$ has more than one neighbor in $N_4$, i.e., $N_5 \subset I$ by Lemma \ref{lemm:structure5.0}.

\begin{lemma}\label{lemm:structure5.1edgeless}
No vertex of $N_5$ has more than one neighbor in $N_4$, i.e., $N_5 = \emptyset$.
\end{lemma}

{\bf Proof.} Suppose to the contrary that $|N(h) \cap N_4| \geq 2$ for $h \in N_5$. As shown in the proof of Lemma \ref{lemm:structure5.0}, there is a vertex $c \in N_3$ such that $c$ sees every vertex of $N(h) \cap N_4$. Thus every edge incident onto $h$ is in a $C_4$ (and thus not in $M$). Then, since $N_5 \subset I$ and since by (\ref{noMedgesN3N4}) no edge between $N_3$ and $N_4$ is in $M$, the edges of such $C_4$'s are not dominated which is a contradiction.
%the only possible way of dominating the edges of such $C_4$'s is that $c=u'_i$ for some $i$ and there is a vertex $d \in N(h) \cap N_4$ with $dh \in M$ but now, there is a $P_8$ with $d,h,a,c$ or $d,h,b,c$ and vertices in $N_1 \cup N_2$ and $x,y$ which is a contradiction.
\qed

\medskip

Thus, from now on, we can assume that $N_5=\emptyset$ and $N_4$ is an independent set.

\begin{lemma}\label{lemm:N4empty}
If $w \in N_4$ and $w' \in N_3$ is a neighbor of $w$ then $w'$ is an $M$-mate $u'_i$ of some $u_i$, and thus, every $w \in N_4$ leads to $xy$-forced $M$-edges.
\end{lemma}

{\bf Proof.} Since we can assume that $N_5=\emptyset$, $N_4$ is an independent set and there is no $M$-edge in $N_3$, edges between $N_3$ and $N_4$ must be dominated by $M$-edges $u_iu'_i$. The only possible way is that every neighbor $w' \in N_3$ of $w \in N_4$ is an $M$-mate $u'_i$ of some $u_i$.
\qed

\medskip

From now on, we can assume that $N_4 = \emptyset$.

\section{A polynomial-time algorithm for DIM on $P_8$-free graphs}

In this section let us describe a polynomial-time algorithm to solve DIM on $P_8$-free graphs.

The main part of the algorithm is simple: For every edge $xy$ in a $P_3$ of $G$ apply the subsequent procedure DIM-with-$xy$, which either returns a proof that $G$ has no d.i.m.\ with $xy$ or returns a minimum (finite) weight d.i.m.\ of $G$ with $xy$ (by the results introduced above). Note that every possible d.i.m.\ $M$ has to be checked whether it is really a d.i.m.; this can be done in linear time for each candidate $M$ (see \cite{BraMos2014}). 

\medskip

{\bf Procedure DIM-with-$xy$}

\medskip

\noindent
{\bf Given:} A connected ($P_8,K_4$,diamond,butterfly)-free $G = (V,E)$ with edge weights, and an edge $xy \in E$ of finite weight which is part of a $P_3$ in $G$.

\noindent
{\bf Task:} Return a proof that $G$ has no d.i.m.\ $M$ with $xy \in M$ (STOP with failure), or return a d.i.m.\ $M$ with $xy \in M$ of finite minimum weight (STOP with success).

\begin{enumerate}

\item Set $M:= \{\{x,y\}\}$. Determine the distance levels $N_i = N_i(xy)$, $1 \le i \le 5$, with respect to $xy$.

\item Check if $N_1$ is an independent set (see condition (\ref{N1subI})) and $N_2$ is the disjoint union of edges and isolated vertices (see condition (\ref{N2M2S2})).
If not, then STOP with failure.

\item For the set $M_2$ of edges in $N_2$, apply the Reduction Step for every edge in $M_2$ correspondingly. Moreover, apply the Reduction Step for each edge $bc$ according to condition (\ref{triangleaN3bcN4}) and then for each edge $u_it_i$ according to Lemma \ref{lemm:structure2} $(v)$.

\item If $N_4 \neq \emptyset$ then, using the results of Subsections \ref{subsec:N4components} and \ref{subsec:N5components} according to the $xy$-forced $M$-edges and the polynomially many cases described in Corollaries \ref{commonindextriangles}, \ref{2N4-1N5-triangleMimplic}, and \ref{1N4-2N5-triangleMimplic}, split the problem into polynomially many such cases. Then, since each such case allows us to finally reduce the problem to the case in which $N_4 = \emptyset$, solve each such case according to the next step and choose a minimum finite weight solution (if such a solution exists).

\item $\{$Now $N_4 = \emptyset.\}$ Apply the approach described in Section \ref{section:N4empty}. Then either return that $G$ has no d.i.m.\ $M$ with $xy \in M$ or return $M$ as a d.i.m.\ of smallest finite weight with $xy \in M$.

\end{enumerate}

\begin{theorem}\label{theo:procedureDIMxy}
Procedure DIM-with-$xy$ is correct and runs in polynomial time.
\end{theorem}

{\bf Proof.} The correctness of the procedure follows from the structural analysis of $P_8$-free graphs with a d.i.m.

The polynomial time bound follows from the fact that Steps 1, 2 can clearly be done in polynomial time, Step 3 can be done in polynomial time since the Reduction Step can be done in polynomial time, Step 4 can be done in polynomial time by the results in Section~\ref{section:nonempty}, and Step 5 can be done in polynomial time as shown in Section \ref{section:N4empty}.
\qed

\medskip

Since a graph $G$ with a d.i.m.\ is $K_4$-free, we can assume that the input graph is $K_4$-free.

\medskip

{\bf Algorithm DIM-$P_8$}

\

\noindent
{\bf Given:} A connected $(P_8,K_4)$-free graph $G = (V,E)$ with edge weights.

\noindent
{\bf Task:} Determine a d.i.m.\ of $G$ of finite minimum weight if one exists or find out that $G$ has no d.i.m.\ of finite weight.

\begin{enumerate}
\item[(a)] Determine the set $F_1$ of all mid-edges of diamonds in $G$, and the set $F_2$ of all peripheral edges of butterflies in $G$. Let $M:=F_1 \cup F_2$. Check whether $M$ is an induced matching in $G$. If not then STOP - $G$ has no d.i.m. Otherwise, check whether $M$ is a dominating edge set of $G$. If yes, we are done. Otherwise apply the Reduction Step for every edge in $F_1 \cup F_2$; without loss of generality, assume that the resulting graph $G'=(V',E')$ is connected (if not, do the next steps for each connected component of $G'$). Let $G:=G'$.

$\{$From now on, $G$ is $(P_8,K_4$, diamond, butterfly)-free.$\}$

\item[(b)] Check whether $G$ has a single edge $uv \in E$ of finite weight which is a d.i.m.\ of $G$. If yes then select such an edge with smallest weight as output and STOP - this is a d.i.m.\ of $G$ of finite minimum weight.

$\{$Otherwise, every d.i.m.\ of $G$ would have at least two edges.$\}$

\item[(c)] For each edge $xy \in E$ of finite weight in a $P_3$ of $G$ carry out procedure DIM-with-$xy$. If DIM-with-$xy$ stops with failure for all edges $xy$ in a $P_3$ of $G$, then STOP - $G$ has no d.i.m. Otherwise, select the best result from all successful applications of the procedure DIM-with-$xy$. If the result does not have finite weight then STOP - $G$ has no d.i.m.\ of finite weight. Otherwise, STOP and return the best result as solution.
\end{enumerate}

\begin{theorem}\label{theo:procedureDIM}
Algorithm DIM-$P_8$ is correct and runs in polynomial time.
\end{theorem}

{\bf Proof.} The correctness of the procedure follows from the structural analysis of $P_8$-free graphs with a d.i.m. In particular: concerning Step (b), one can easily verify that if $G$ has a d.i.m.\ of one edge, then $G$ has no d.i.m.\ with more than one edge; concerning Step (c), one can refer to Observation \ref{obse:xy-in-P3}. The time bound follows from the fact that Step (a) can be done in polynomial time (in particular the Reduction Step can be done in polynomial time), Step (b) can be done in polynomial time, and Step (c) can be done in polynomial time by Theorem \ref{theo:procedureDIMxy}.
\qed

\medskip

{\bf Acknowledgments.} The authors gratefully thank three anonymous reviewers for their helpful comments. 
The second author would like to witness that he just tries to pray a lot and is not able to do anything without that.

\begin{footnotesize}

\end{footnotesize}

\end{document}